\documentclass{PoS}
\newcommand{\be}{\begin{eqnarray}}
 \newcommand{\ee}{\end{eqnarray}}
 \newcommand{\nee}{\nonumber\end{eqnarray}}
 \newcommand{\nn}{\nonumber\\}

\def\b               {\beta}
\def\d               {\delta}

\def\t               {\theta}

\def\x               {\chi}

\def\ti              {\tilde}

\def\snu             {\ti\nu}  
\def\stau            {\ti\tau}
\def\st              {\ti t}
\def\sb              {\ti b}
\def\ch              {\ti\x^\pm}

\def\nt              {\ti\x^0}
\def\sg              {\ti g}

\def\dcp             {\d^{CP}_{\nu\tau}}

\newcommand{\msnu}     {m_{\snu}}
\newcommand{\mstau}[1] {m_{\stau_{#1}}}
\newcommand{\mst}[1]   {\msnu}
\newcommand{\msb}[1]   {m_{\stau_{#1}}}

\newcommand{\mhp}      {m_{H^+}}
\newcommand{\mh}       {m_{H^\pm}}

\renewcommand{\Re}{{{\cal R}e}}

\title{CP violation in charged Higgs boson decays in the MSSM}

\ShortTitle{CP violation in charged Higgs boson decays}

\author{\speaker{Ekaterina Christova}\\
        Institute for Nuclear Research and Nuclear Energy of BAS, Sofia, Bulgaria\\
        E-mail: \email{echristo@inrne.bas.bg}}

\author{Helmut Eberl\\
        Institut f\"ur Hochenergiephysik der \"OAW, A-1050 Vienna,
        Austria\\
        E-mail: \email{helmut@hephy.oeaw.ac.at}}

\author{Elena Ginina\thanks{On leave of absence from the Institute for
             Nuclear Research and Nuclear Energy of BAS, Sofia, Bulgaria}\\
        Institut f\"ur Hochenergiephysik der \"OAW, A-1050 Vienna,
        Austria\\
        E-mail: \email{eginina@hephy.oeaw.ac.at}}

\abstract{CP violation in $H^\pm$ decays into the three possible
decay modes into ordinary particles, 1) $H^\pm \to tb$, 2) $H^\pm
\to \nu \tau$ and 3) $H^\pm \to W^\pm h^0$ is considered. Analytic
expressions and numerical results for the CP~violating decay rate
asymmetries in the MSSM are obtained. Increasing $\tan\beta$  the
asymmetries for the fermionic decays, $H^\pm \to tb$ and $H^\pm
\to \nu \tau$, decrease and it increases for $H^\pm \to W^\pm
h^0$. The asymmetry of $H^\pm \to tb$ is most sensitive to the
phase of $A_t$ and can go up to 20\%, the asymmetries of  2) and
3) depend mainly on the phases of $A_\tau$ and $M_1$. The
asymmetry of 2) is smaller than 0.5\% and of 3) can reach up to
2\%.}

\FullConference{Prospects for Charged Higgs Discovery at Colliders\\
         September 16-19 2008\\
         Uppsala, Sweden}

\begin{document}

\section{Introduction}

If a charged Higgs boson is discovered at LHC, which we believe
very much, or at a possible future International Linear Collider (ILC) or at CLIC,
it would be ultimately a signal for Physics beyond the Standard Model (SM).
The next question would be which Physics beyond the SM is it --
almost all extensions of the SM enlarge the Higgs sector of the SM
and inevitably predict the existence of a charged Higgs.
The effects of CP violation is a possible tool to disentangle
the different charged Higgs bosons. Nearly all extensions of the
SM contain additional sources of CP violation.

In this note we study CP violation in the Minimal Supersymmetric
Standard Model (MSSM) with complex couplings. We consider the
processes of $H^\pm$-decays into ordinary particles -- these are
the decays
\be
H^\pm \to tb,\qquad H^\pm \to \nu \tau\qquad {\rm and}\quad H^\pm \to
W^\pm h^0,
\ee
where $h^0$ is the lightest neutral Higgs boson.

The CP violating asymmetries that we consider are the decay rate
asymmetries
\be
\delta^{CP}_f=\frac{\Gamma (H^+ \to f)- \Gamma (H^- \to \bar f)}
{\Gamma (H^+ \to f)+\Gamma (H^- \to \bar f)}\label{asymm}
\ee
where $f$ stands for $tb$, $\nu \tau$ and $W h^0$, respectively.
At tree level the partial decay rates are always equal and there is no CP violation.
$\delta^{CP}_f$ is a loop induced effect, in our case these are loops with SUSY particles.
However, a CP violating phase and loop corrections are not enough --
 for $\delta^{CP}_f \neq 0$ the loop integrals must have absorptive parts,
i.e. $\delta^{CP}_f$ is a threshold effect -- for a non-zero value of $\delta^{CP}_f$
at least one decay channel of $H^\pm$
 into SUSY particles should be open.

In the SM the only source of CP violation is the CKM CP phase. In the MSSM, in addition to it,
new phases appear -- these are the phase of the higgsino mass parameter $\mu = \vert \mu\vert e^{i\phi_\mu}$,
the phases of the gaugino masses $M_i=\vert M_i\vert e^{i\phi_i}$, $i=1,2,3$
and the phases of the trilinear couplings $A_f=\vert A_f\vert e^{i\phi_f}$.
Of these the phase $\phi_\mu$ is strongly constrained by measurements of the neutron and
eletron EDM: $\phi_\mu\leq 10^{-2}$. The phases of the trilinear couplings $A_f$
always occur as $A_f\,m_f$, with $m_f$ the corresponding fermion mass.
They are practically only important for the fermions of the 3-rd generation only. Thus, the phases most
relevant to our study are
$\phi_t$, $\phi_b$ and $\phi_\tau$ -- the phases of $A_t$, $A_b$ and $A_\tau$.
We also allow for a non-zero phase $\phi_1$ of $M_1$, imposing the GUT relation
only for the absolute values, $|M_1|=\frac{5}{3}\tan\t_W\,|M_2|$, because the phase of $M_2$ can be rotated
away and is not physical.

In refs.~\cite{we1}--\cite{we4}  these decay rate asymmetries were studied. Here we give
a short review of these papers.

\section{$H^\pm \to tb$ decay}

\begin{figure}[h!]
 \begin{center}
 \mbox{\resizebox{!}{4.8cm}{\includegraphics{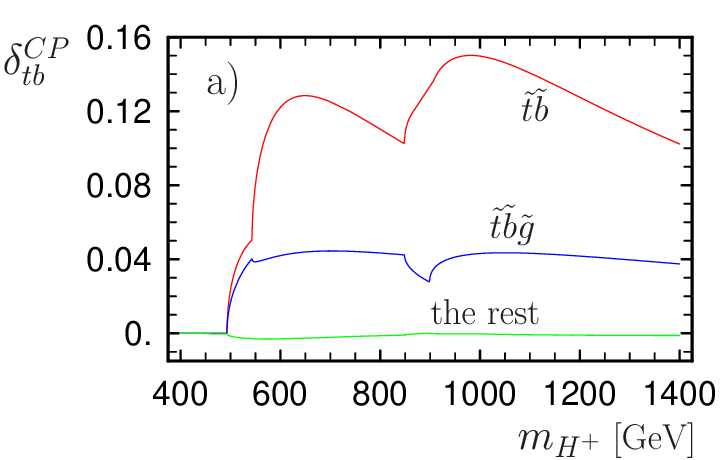}}
 \hspace*{-0.2cm}
 \resizebox{!}{4.8cm}{\includegraphics{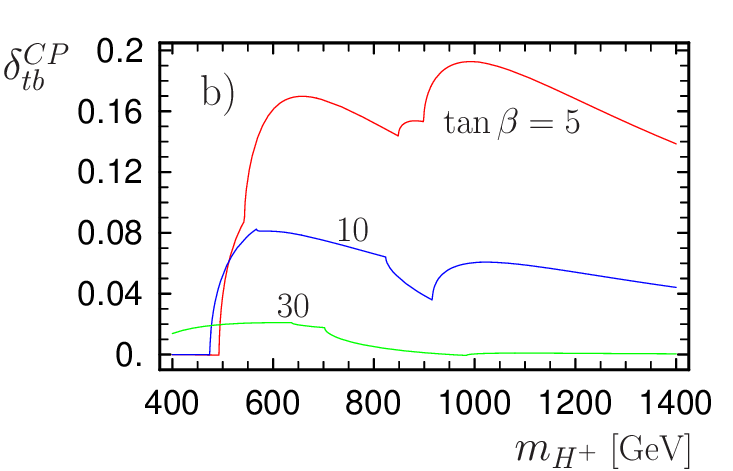}}} \hfil
 \vspace*{-10mm}
 \end{center}
 \caption{ a) the different contributions to
 $\delta^{CP}_{t b}$,\, $\tan\beta =5$,
 and
 b) $\delta^{CP}_{t b}$ as a function of $m_{H^+}$, $\tan\beta =5,\, 10$ and $30$.
 For both figures $\phi_{A_t}=\pi/2$, \,
 $\phi_{A_b}=\phi_\mu = \phi_3 = 0$}
 \label{fig1}
\end{figure}

The matrix elements for $H^\pm\to t\bar b$  decays, including loop corrections, can be written as
\be
{\cal M}_{H^+}&=&\bar u(p_t)\left[ Y^+_b\,P_R\,+Y^+_t\,P_L\,\right]v(-p_{\bar b}),\nn
{\cal M}_{H^-}&=&\bar u(p_b)\left[ Y^-_b\,P_R\,+Y^-_t\,P_L\,\right]v(-p_{\bar t}),
\ee
where $P_{R,L}=(1\pm \gamma_5)/2$ and the form factors  are
\be
Y_i^\pm = y_i\pm \delta Y^\pm_i\,,\qquad \delta Y_i^\pm=\delta Y_i^{inv} \pm \delta Y_i^{CP}\,,\qquad
i=t,b\, ,
\ee
$y_i$ are the tree level Yukawa couplings.
The loop induced form factors  $\delta Y_i^\pm$  have CP-invariant   and CP violating
contributions. The CP violating contributions $\delta Y_i^{CP}$  distinguish  the form
factors of $H^+$  and  $H^-$.
Both $\delta Y_i^{inv}$ and $\delta Y_i^{CP}$ have real and imaginary (absorptive) parts.
In the decay rate asymmetries always the absorptive parts contribute.
This can be easily understood following the simple explanation: for having CP violation we need a CP phase,
the other phase that we need in order to have a real decay width could come only
from absorptive parts of the loop integrals. For the decay rate asymmetry $\delta^{CP}_{t b}$ we obtain~\cite{we1}
\be
\delta^{CP}_{t b}=\frac{2(m_{H^+}^2-m_t^2-m_b^2)(y_t\Re  \delta Y_t^{CP}+y_b\Re  \delta Y_b^{CP})-4m_tm_b
(y_t\Re \delta Y_b^{CP}+y_b\Re  \delta Y_t^{CP})}{(m_{H^+}^2-m_t^2-m_b^2)(y_t^2+y_b^2)-4m_tm_b\,y_ty_b}.
\ee
At one loop there are two types of SUSY corrections:
in  the $H^\pm tb$-vertex and self-energy corrections on the $H^\pm$-line.
They were calculated in ~\cite{we1} and analytic expressions for the results are given therein.
The performed numerical analysis~\cite{we1,we2} showed that the contributions from the self-energy loop
with $\st \sb$ give the main contribution to $\delta^{CP}_{t b}$, also the $\sg\st\sb$-vertex corrections
can give a relevant contribution,  the contributions with $\ch$, $\nt$ and $\ti\tau$ are totally negligible.
The  relative importance of the different diagrams is shown on Fig.\ref{fig1}a).
This suggests that  $\delta^{CP}_{t b}$ will be  sensitive to the phases of $A_t$ and $A_b$ only.
As the contribution of $A_t$ is enhanced by the large mass of the $t$-quark mass, $m_t=178 $ GeV,
the dependence on $A_b$ should be much weaker. Our numerical analysis showed that there is
no sensitivity to the phase of $A_b$.

We have studied the dependence of $\delta^{CP}_{t b}$ on $m_{H^+}$ and the phases $\phi_t$ and $\phi_b$ for
different values of $\tan\beta$.
In order not to vary too many parameters, we fix:
$M_2=300~{\rm GeV}, M_3=745~{\rm GeV}, M_{\tilde U}= M_{\tilde Q}= M_{\tilde D}= M_E= M_L =350~ {\rm GeV},
\mu=-700~{\rm GeV}, |A_t| = |A_b| = |A_\tau|=700~{\rm GeV}$.
The relevant sparticle masses for this choices are given explicitly
in Table~\ref{table1},

Fig~\ref{fig1}b) shows $\delta^{CP}_{t b}$ as a function of $m_{H^+}$ for different values of $\tan\beta$.
For $\mhp < \mst{1}+\msb{1}$, $\d^{CP}$ is very small,
${\cal O}(10^{-3})$ or smaller. However, once the $H^+\to\st\bar{\sb}$ channel is open,
$\d^{CP}$ can go up to roughly 20\%. All four thresholds of $H^+\to\st_i \bar{\sb_i}$
are clearly visible in the figure.

The considered decay mode $H^\pm \to t b$ will be traced by the decay products of the $t$-quark. Because of its
large mass, the $t$-quark will decay keeping its momentum and polarization. In ref.~\cite{we2} the CP violating angular
and energy asymmetries of the $t$-decay products are considered as well.

\newcommand{\myx}{@{\hspace{3mm}}}
\begin{table}
\begin{center}
\begin{tabular}{|c||c\myx c\myx c\myx c|c\myx c|c\myx c|c\myx c|c\myx c|c|}
\hline
   $\tan\beta$
  & $m_{{\tilde \chi}^0_1}$ & $m_{{\tilde \chi}^0_2}$ & $m_{{\tilde \chi}^0_3}$ & $m_{{\tilde \chi}^0_4}$ &
   $m_{{\tilde \chi}^+_1}$ & $m_{{\tilde \chi}^+_2}$ &
   $m_{{\tilde t}_1}$ &  $m_{{\tilde t}_2}$  &  $m_{{\tilde b}_1}$  &  $m_{{\tilde b}_2}$  &
   $m_{{\tilde \tau}_1}$ & $m_{{\tilde \tau}_2}$ & $m_{{\tilde \nu}}$ \\
\hline
 5 & 142 & 300 & 706 & 706 & 300 & 709 & 166 & 522 & 327 & 377 & 344 & 362 & 344 \\
 30 & 141 & 296 & 705 & 709 & 296 & 711 & 172 & 519 & 183 & 464 & 295 & 402 & 344\\
\hline
\end{tabular}
\end{center}
\vspace*{-5mm}
\caption{Sparticles masses (in GeV) for the given parameter together with
$\phi_{A_t}=\phi_{A_b}=\pi/2$ and
$\phi_\mu$ = 0.
\label{table1}}
\end{table}

\section{$H^\pm \to \nu\tau^\pm$ decay}

In the previous section we showed that large phases of $A_{t}$  can lead to a large CP-violating asymmetry
$\d^{CP}_{tb}$ in $H^\pm \to tb$,  up to
of 15\,--\,20\% for $\mh > m_{\tilde{t}} + m_{\tilde{b}}$.
In this section we consider the lepton decay channels of the charged
Higgs bosons, $H^+\to\tau^+\nu_\tau$ and $H^-\to\tau^-\bar\nu_\tau$ and
calculate the CP-violating asymmetry $\dcp$ at the one-loop level.

The decay  $H^\pm\to\tau\nu$ may be important for relatively low masses
of $H^\pm$ when the decay $H^\pm \to tb$ is not allowed kinematically. This implies that,
as it is always the absorptive parts of the loop integrals that contribute,
the loops with $\st$ and $\sb$ will not contribute to $\dcp$ in this region.
Thus,
the only relevant phase from the trilinear couplings will be the phase $\phi_\tau$.
As we also allow for the gaugino mass parameter $M_1$ to be complex, a non-zero value of
$\dcp$ would imply non-zero phases $\phi_\tau$ and/or $\phi_1$.

The theoretical consideration is quite similar to the decay $H^\pm \to t b$, but as $m_\nu =0$
there is only one form factor in the matrix element:
\be
Y_\tau^\pm = y_\tau +\delta Y_\tau^\pm ,\quad \delta Y_{\tau}^\pm =\delta Y_{\tau}^{inv}\pm \delta Y_{\tau}^{CP}
\ee
where $y_\tau$ is the tree level coupling. Both the CP-invariant and the
CP-violating contributions have real and imaginary parts and $\dcp$ is expressed in terms
of the imaginary part in the simple form:
\be
  \dcp =
  \frac{2\Re\,\d Y_\tau^{CP}}{y_\tau\,+\,2\,\Re\,\d Y_\tau^{inv}}
  \simeq \frac{2\Re\,\d Y_\tau^{CP}}{y_\tau} \,.
\label{eq:DCP}
\ee
The loops that will contribute are: self-energy loops with $\stau \ti\nu$ and $\ti\chi^+\ti\chi^0$, and vertex graphs with
$\stau\ti\nu\ti\chi^0$, $\ti\chi^+\ti\chi^0\ti\nu$ and $\ti\chi^-\ti\chi^0\ti\tau$. The explicit expressions
for $\d Y_\tau^{CP}$ from the different loop diagrams,
together with the masses and couplings of staus and sneutrinos,
are given in~\cite{we3}.

For the numerical analysis we fix $M_2=200~{\rm GeV},~ \mu=300~{\rm GeV},~
M_{\ti E} = M_{\ti L} - 5~{\rm GeV},~
|A_\tau| = 400~{\rm GeV}.$

\begin{table}[h!]\center
\begin{tabular}{|c|r|r|r|r|}
\hline
   $\tan\beta$ & $M_{\ti L}$ & $\msnu$ & $\mstau{2}$ & $\t_{\stau}$ \\
\hline
   5 & 138 & 123 & 150 & $56^o$ \\
  10 & 147 & 132 & 166 & $50^o$ \\
  30 & 180 & 168 & 221 & $47^o$ \\
\hline
\end{tabular}
\caption{Parameters and slepton masses in [GeV] used in the analysis for $\delta^{CP}_{\nu\tau}$,
$\mstau{1}=135$~GeV.}
\label{table2}
\end{table}

\begin{figure}[h!]
 \begin{center}
 \mbox{\resizebox{!}{6cm}{\includegraphics{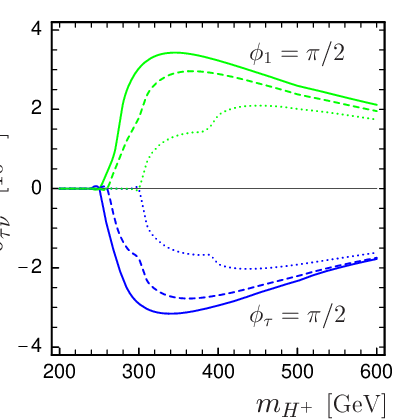}}} \hfil
  \end{center}
  \vspace*{-6mm}
  \caption {$\d^{CP}_{\nu \tau}$ as a function of $\phi_\tau$ for
  $\mh=350$~GeV and $\tan\b=5$.
  The full, dashed, and dotted lines are for
  $\phi_1=0$, $\pi/4$, and $\pi/2$, respectively.}
  \label{fig2}
\end{figure}

Fig.~\ref{fig2} shows $\dcp$ as a function of $\mh$ for the two cases:
$\phi_\tau=\pi/2$, $\phi_1 = 0$ and $\phi_\tau=0$, $\phi_1 = \pi/2$
and  $\tan\b=5$, 10, and 30. The corresponding values for $\msnu$,
$\mstau{2}$ and $\t_{\stau}$ are listed in Table\ref{table2}. In our
analysis $|\dcp|$ goes up to $\sim 3.5\times 10^{-3}$
 and it is interesting to note that maximal $\phi_\tau$ and
maximal $\phi_1$ lead to very similar values of $\dcp$
but with opposite signs.
However, if both phases are maximal, i.e.
$\phi_\tau\sim\phi_1\sim\pi/2$ or $3\pi/2$, they compensate
each other and $\dcp$ practically vanishes.

\section{$H^\pm \to W^\pm h^0$ decay}

In this section we consider
the decay rate asymmetry $ \delta^{CP}_{W h^0}$ of $H^\pm \to W^\pm h^0$ decay.
Though the final
state $h^0$ is not observed yet,  $m_{h^0}$ is not an unknown
parameter -- once $m_{H^+}$ and $\tan\beta$ are fixed, the
SUSY structure of the theory determines uniquely both $m_{h^0}$
and the branching ratio (BR).
Previously this asymmetry was
considered in the two-Higgs doublet model~\cite{Lavoura}.

Respecting the experimental lower bound from LEP and the theoretical upper bound,
including radiative corrections,
we consider $m_{h^0}$ in the range 96 $ \leq m_{h^0}\leq$ 130 GeV.
In order to keep the value of BR($H^+\to W^+h^0)$ at the level of a few percent,
we consider low $m_{H^+}$,  200 $\leq m_{H^+}\leq 600$ GeV and low $\tan\beta$,
$3 \leq \tan\beta \leq 9$ ($\tan\beta \leq 3$ being excluded from
the Higgs searches at LEP).

The matrix elements of $H^+\to W^\pm h^0$ is expressed in terms of one form factor only:
 \be
M_{H^\pm}=ig\varepsilon_{\alpha}^{\lambda}(p_W)p_h^{\alpha}Y^\pm,\quad Y^\pm=y+\delta Y^\pm ,\quad
\delta
Y^{\pm}=\delta Y^{inv} \pm \delta Y^{CP}
\ee%
where $y=\cos (\alpha - \beta)$ is the tree level coupling. For $\delta^{CP}_{W h^0}$ we obtain~\cite{we4}:
\be
 \delta^{CP}_{W h^0}\simeq
 {2 \Re (\delta Y^{CP})\over y }.
\ee
As in the previous section, we assume that the squarks, as suggested by most SUSY models,  are heavier
and thus will not contribute in the considered range of $m_{H^+}$.

In accordance with this, there are two types of diagrams that will contribute: with  $\ti\nu,\ti\tau$ and with
$\ti\chi^\pm$ and $\ti\chi^0$ in the loops. The explicit expressions were obtained in \cite{we4}.
This implies the sensitivity  of  $\delta^{CP}_{W h}$ to the phases
$\phi_\tau$ and $\phi_1$.  The numerical analysis was performed for
\be
  M_2=250~{\rm GeV},~
     M_{ E} = M_{ L} - 5~{\rm GeV},M_L=120~{\rm GeV},~
|A_\tau| = 500~{\rm GeV},~
     \vert \mu \vert=150~{\rm GeV.}~\label{leppars}
        \ee
        and the GUT relation for the absolute values of
$M_1$ and $M_2$ was assumed.

 The numerical analysis showed that the values for $\delta^{CP}_{W h^0}$ are typically about
~ $10^{-2}\div10^{-3}$, the main contributions being from
$\tilde\nu$ and $\tilde\tau$ for $m_{H^+} < $ 300 GeV, and from
$\tilde\chi^+$ and $\tilde\chi^0$ for $m_{H^+} \geq $ 300 GeV.
 The dependence on different values of $\tan\beta$ was examined~\cite{we4}.

\section{Summary}

Discussing CP violation in the decay widths, we
must keep in mind the branching ratios of the relevant decay
modes. The BR of $H^+\to \nu\tau^+$ is dominant below
the $t \bar b$ threshold. This determines the sensitivity of
$\delta^{CP}_{\nu\tau}$ to the phases
$\phi_\tau$ and $\phi_1$ of $A_\tau$ and $M_1$, respectively.
The decay rate asymmetry remains always below 0.5\%.

 If  $m_{H^+}$ is large enough and the $t\bar b$-threshold  is open,
$H^+\to t\bar b$ will  dominate
and $\delta^{CP}_{t b}$ will be important. Due to the large top
Yukawa coupling and the fact that $\delta^{CP}_{t b}$ goes down
for large $\tan\beta$, for all values of $\tan\beta$ it is most
sensitive to the phase of $A_t$. For large $m_{H^+}$ and a
relatively light gluino, $m_{\tilde g} \sim 400$~GeV, and light
stops and sbottoms, $m_{\tilde t} = 166$~GeV and  $m_{\tilde b} =
327$~GeV, $\delta^{CP}_{t b}$ can go up to $\sim$ 20\%.

The decay rate asymmetry $\delta^{CP}_{W h^0}$ can be of the order of few percents
if both $m_{H^+}$ and $\tan\beta$ are  small and
will be sensitive to $\phi_\tau$ and $\phi_1$. The BR of $H^+\to W^+
h^0$ can go up to 10\%.

In principle, these asymmetries could be directly measured at an ILC or at CLIC if $\sqrt s > 2 m_{H^+}$.
But after doing a more detailed estimation, in all three cases a
higher luminosity would be necessary to observe CP violation in
these decays.

For LHC one must take into account CP violation in the production of $H^\pm$ as
well, see the contribution \cite{elena_ginina} within these
proceedings.

\section*{Acknowledgements}
%

The authors acknowledge support from EU under the MRTN-CT-2006-035505 network
programme. This work is supported by the "Fonds zur F\"orderung der
wissenschaftlichen Forschung" of Austria, project No. P18959-N16.

\end{document}